\documentclass[prl,showpacs,showkeys,twocolumn,floatfix,superscriptaddress]{revtex4}

\usepackage{graphicx}\usepackage{units}\usepackage{epsfig}\usepackage{xspace}\usepackage[american]{babel}\usepackage{color}

\begin{document} 

\title{Collective oscillations of dipolar Bose-Einstein condensates  and \\
accurate comparison  between contact and dipolar interaction}

\author{S.\ Giovanazzi} 	\email{stevbolz@yahoo.it}
\affiliation{5.\ Physikalisches Institut, Universit\"at Stuttgart, Pfaffenwaldring 57, 70550 Stuttgart, Germany} 
\author{L.\ Santos} 		
\affiliation{Institut f\"ur Theoretische Physik III, Universit\"at Stuttgart, Pfaffenwaldring 57, 70550 Stuttgart, Germany}
\affiliation{Institut f\"ur Theoretische Physik, Universit\"at Hannover, Appelstr. 2, 30167 Hannover, Germany.}
\author{T.\ Pfau}		
\affiliation{5.\ Physikalisches Institut, Universit\"at Stuttgart, Pfaffenwaldring 57, 70550 Stuttgart, Germany}
\date{\today}

\begin{abstract} 
We propose a scheme for the measurement of the s-wave scattering length $a$ of an atom or molecule with significant dipole-dipole interaction with an accuracy at the percent level. The frequencies of the collective oscillations of a Bose-Einstein condensate are shifted by the magnetic dipole interaction. The shift is polarization dependent and proportional to the ratio $\varepsilon_{\mathrm{dd}} $ of dipolar and s-wave coupling constants.
Measuring the differences in the frequencies for different polarization we can extract the value of $\varepsilon_{\mathrm{dd}} $ and thus measure $a$.  
We calculate the frequency shifts for a large variety of non-axisymmetric harmonic traps in the Thomas-Fermi limit and find optimal trapping geometries to maximize the shifts. 
\end{abstract} 
\maketitle

Recent research on polar molecules \cite{polarmolecules}, Rydberg atoms \cite{rydbergatoms}, laser-induced dipoles \cite{laser-induced}, dipolar spinor condensate \cite{spinor} and Bose-Einstein condensates (BECs) of polarized chromium \cite{chromium,Giovanazzi:2006ex} is attracting a growing attention towards the exciting new physics of dipolar gases. 
Dipolar gases have intricate stability diagrams and ground state properties \cite{ground,Goral:2002,ODell:2004ex,Eberlein:2005ex}, including new quantum phases in optical lattices \cite{Goral:2002a} and roton dispersion relations \cite{rotons}.

In this paper, we present exact semi-analytical results for the frequencies of the collective condensate oscillations in non-axisymmetric traps. We use recent analytical results  \cite{ODell:2004ex,Eberlein:2005ex,Giovanazzi:2006ex} obtained in the experimentally relevant Thomas-Fermi (TF) regime of a large number of atoms or molecules $N$. Previous studies of collective excitations were confined to the case of axisymmetric traps \cite{excitations,Goral:2002,LiYou:2002pr}.

As an application of our results, we discuss a new method for the accurate determination of the s-wave scattering length $a$ at the percent level. This possibility is entirely due to the dipolar interaction and its high degree of manipulability.
Measurements of $a$ at 1$\%$ accuracy level near a Feshbach resonance may be used to detect ultra-small variations of the fundamental constants \cite{Chin:2006en}.

The density $n$ of a dipolar BEC and its superfluid velocity $\vec{v}$ obey the continuity and Euler equations given by 
\begin{eqnarray} {\partial n \over \partial t} &=& - {\vec{ \nabla}} \cdot (n \vec{v}) \label{continuity}\\
m{\partial {\vec{ v}} \over \partial t} &=&-  {\vec{ \nabla}} \left(\frac { m v^2}{2} + V_{\mathrm{ho}} + V_{\mathrm{mf}}\right) \label{euler} \end{eqnarray}
where $V_{\rm ho}=(m/2)\left(\omega_x^2 x^2+\omega_y^2  y^2+\omega_z^2 z^2\right)$ is the trapping potential and $V_{\rm mf}$ is the mean field potential given by
\begin{equation} \label{mf} V_{\rm mf} = \frac{4 \pi \hbar^2 \,a }{m} \, n(\vec{r},t) +  \int d^3r' V_{\rm dd}(\vec{r}-\vec{r}\,') n(\vec{r}\,',t) \end{equation}
where $m$ is the mass of the dipolar particle. $V_{dd}$ is the anisotropic and long-range dipolar interaction given by 
\begin{equation}\label{udd} 
V_{\rm dd}(\vec{r})=\frac{ C_{\mathrm{dd}}}{4 \pi r^3} \left( 1- \frac{ 3 (\hat{e}_\mu\vec{r})^2}{r^2}\right) 
\end{equation}
where $C_{\mathrm{dd}}=\mu_0 \mu^2_{\rm m}$ for magnetic particles of moment $\mu_m$ and $C_{\mathrm{dd}}={\rm d}^2_{\rm e}/\epsilon_0$ for particles with an electric dipole moment ${\rm d}_e$.
The properties of dipolar BEC depend additionally on the dipole orientation $\hat{e}_\mu$ and on the dimensionless ratio between the dipolar coupling $C_{\mathrm{dd}}$ and $a$
\begin{eqnarray} 
\varepsilon_{\mathrm{dd}} = \frac{C_{\mathrm{dd}}}{12 \pi \hbar^2 a} \label{epsilon} 
\end{eqnarray}

An {\it exact class of solutions} of the TF hydrodynamic equations (\ref{continuity}) and (\ref{euler}) has been obtained in \cite{ODell:2004ex,Eberlein:2005ex,Giovanazzi:2006ex}. The existence of this class of solutions for harmonic traps is due to the harmonic nature of the external and self-consistent potentials and the Bernoulli term. The density has the ``usual'' form of an {\it inverted parabola} given by 
\begin{eqnarray}
n(\vec{r},t) &=& \frac{15 N }{8\pi  R_x R_y R_z} \left[1-\frac{x^2}{R_x^2}-\frac{y^2}{R_y^2} - \frac{z^2}{R_z^2}\right] \label{parabola} 
\end{eqnarray}
The velocity field $\vec{v}(\vec{r},t)=  (1/2) \vec{\nabla} \left[ \alpha_x x^{2} + \alpha_y y^{2}+ \alpha_z z^{2} \right]$ depend on  the condensate radii $R_{j}$ with $j=x,y,z$ through the relation $\alpha_{j}= \partial_{t} \,\log(R_{j}) $. 
The $R_{j}$ are obtained by solving the Newton equation 
\begin{equation}\label{ne}
  \ddot{R_{j}} = -   \partial_{j} u \left( R_x,R_y,R_z\right)
\end{equation}
where $\partial_{j}$ denotes the partial derivative $\partial/\partial_{R_{j}}$. 
The reduced potential landscape $u$ is the sum of  $u_{\mathrm{ho}}=(1/2) \left(\omega_x^2R_x^2+\omega_y^2R_y^2 + \omega_z^2R_z^2\right) $ from the harmonic trap, $ u_{s} =\left( 15 N \hbar^2 a /  m^{2} \right) / R_x R_{y} R_z$ from the s-wave pseudopotential and
$ u_{dd}= - \varepsilon_{\mathrm{dd}} \,f(\kappa_{1},\kappa_{2}) u_{s} $ from the dipole-dipole interaction. 
For dipoles aligned with the $\hat{x}$ axes the condensate aspect ratios $\kappa_{1}, \kappa_{2}$ are given by $\kappa_{1}=R_y/R_x$ and $\kappa_{2}=R_z/R_x$. The symmetric function $f$ with values in the interval $(-2,1)$ is given by
\begin{eqnarray}\label{ffunction} 
f(\kappa_{yx},\kappa_{zx})&=&1+3\kappa_{yx}\kappa_{zx}  {\mathrm{E}(\varphi\setminus\alpha) -\mathrm{F}(\varphi\setminus\alpha) \over (1-\kappa_{zx}^{2})\sqrt{1-\kappa_{yx}^{2}}} 
\end{eqnarray}
with $ \sin\varphi=(1-\kappa_{yx}^{2})^{1/2}$ and $\sin^{2}\alpha=(1-\kappa_{zx}^{2})/(1-\kappa_{yx}^{2}) $.
$\mathrm{F}(\varphi\setminus\alpha)$  and $\mathrm{E}(\varphi\setminus\alpha)$ are the incomplete elliptic integrals of the first and second kinds (see \cite{Giovanazzi:2006ex} for more details). 
\begin{figure}[b] 
\includegraphics[width=89mm]{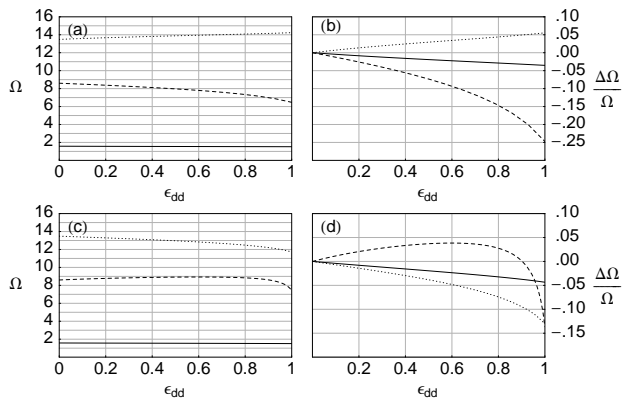}
\vspace{-0.4cm} 
\caption{\label{figfre}
Solid, dashed and dotted lines represent in (a) and (c) the three collective frequencies  $\Omega_{3}$, $\Omega_{2}$ and $\Omega_{1}$ (in units of $\omega_{z}$) and in (b) and (d) the corresponding relative variations $\Delta\Omega/\Omega$ as function of $\varepsilon_{\mathrm{dd}} $ in the TF interval $(0,1)$. The magnetic dipoles are polarized along the $x$-axes in (a) and (b) and along the $y$-axes in (c) and (d). The trap frequencies are 
$\omega_{x} = 2 \pi 942$ Hz, $\omega_{y} = 2 \pi 712$ Hz and  $\omega_{z} = 2 \pi 128$ Hz.  }
\end{figure}

The equations for the small oscillations around equilibrium are obtained expanding the Newton equation (\ref{ne}) for the condensate radii $R_{j}$ around the equilibrium values $R_{j}^{e}$ for small displacements $\delta R_{j}=R_{j}-R_{j}^{e}$
\begin{equation}\label{nel}
 \ddot{\delta  R_{j}} = - u''_{j m} \,\delta R_{m}
\end{equation}
where 
\begin{equation}\label{nel2}
u''_{l m}=\partial_{l m} \,u\vert_{R_i=R^{e}_i}
\end{equation}
is the matrix of the second derivatives of $u$ with respect to the condensate radii $R_{l}$ and $R_{m}$ evaluated on the equilibrium values $R_{j}^{e}$.
The frequencies $\Omega_{k}$ of the small oscillations around equilibrium are simply given by the square roots of the eigenvalues of $u''_{l m}$. The index $k$ of the normal modes are chosen so that $\Omega_{3}\le\Omega_{2}\le\Omega_{1}$.

Fig. 1 shows an example of the dipole-induced shifts on the frequency of the collective modes as a function of $\varepsilon_{\mathrm{dd}}$ between $0$ and $1$ for a possible experimental relevant anisotropic trap. $\varepsilon_{\mathrm{dd}}$ can be varied by tuning the scattering length via a Feshbach resonance (see \cite{Werner:2005} for the case of chromium).
Fig. 1 shows how for sufficiently low values of $\varepsilon_{\mathrm{dd}}$ the $\Omega_{k}$ depend almost linearly on $\varepsilon_{\mathrm{dd}}$. In the following, we shall employ this fact to obtain simplified expressions which allow for a quick and simple estimation of the dipole-induced shifts.

In the absence of dipole-dipole interaction the unperturbed matrix $u''_{l m}$  is given by the simple expression
\begin{equation}\label{unpert}
u''_{l m}\vert_{0}= 2\, \omega_{l}^{2}\,\delta_{l  m}+\omega_{l}\,\omega_{m}
\end{equation}
where $\vert_{0}$ denotes the evaluation at $\varepsilon_{\mathrm{dd}} =0$.
The eigenvalues and eigenvectors  of (\ref{unpert}) can be calculated analytically but their expressions are too long to be reported here.
To first order in $\varepsilon_{\mathrm{dd}} $ the variation of the matrix  $u''_{l m}$  is given by
\begin{eqnarray} \label{nel20}
\Delta u''_{l m}&=&u''_{l m}- u''_{l m}\vert_{0} \nonumber\\&=&
\partial_{l m n} u\vert_{0}\,\Delta R_{n}^{e}
+\partial_{l m}\,u_{dd}
\end{eqnarray}
where $\partial_{l m}\,u_{dd}$ is the dipolar contribution to $u''_{l m}$ evaluated on the unperturbed equilibrium configuration $R_{n}^{e}\vert_{0}$  and 
\begin{eqnarray} \label{nel20}
(R_{l}^{e}\,\partial_{i j l} u)\vert_{0}=&
 -& 2\, \omega_{l}^{2}\,\delta_{i j l}- 2\, \omega_{j}^{2}\,\delta_{i j }
 \nonumber\\&-&\omega_{l}\,\omega_{i}\,\delta_{j l}
-\omega_{l}\,\omega_{j}\,\delta_{i l}-\omega_{i}\,\omega_{j}
\end{eqnarray}
is part of the contribution $\partial_{l m n} u\vert_{0}\,\Delta R_{n}^{e}$ due to the dipolar displacement of the equilibrium $\Delta R_{n}^{e}\vert_{0}=R_{n}^{e}-R_{n}^{e}\vert_{0}$ given by  
\cite{Giovanazzi:2006ex}
\begin{figure*} [h!] 
\includegraphics[width=150mm]{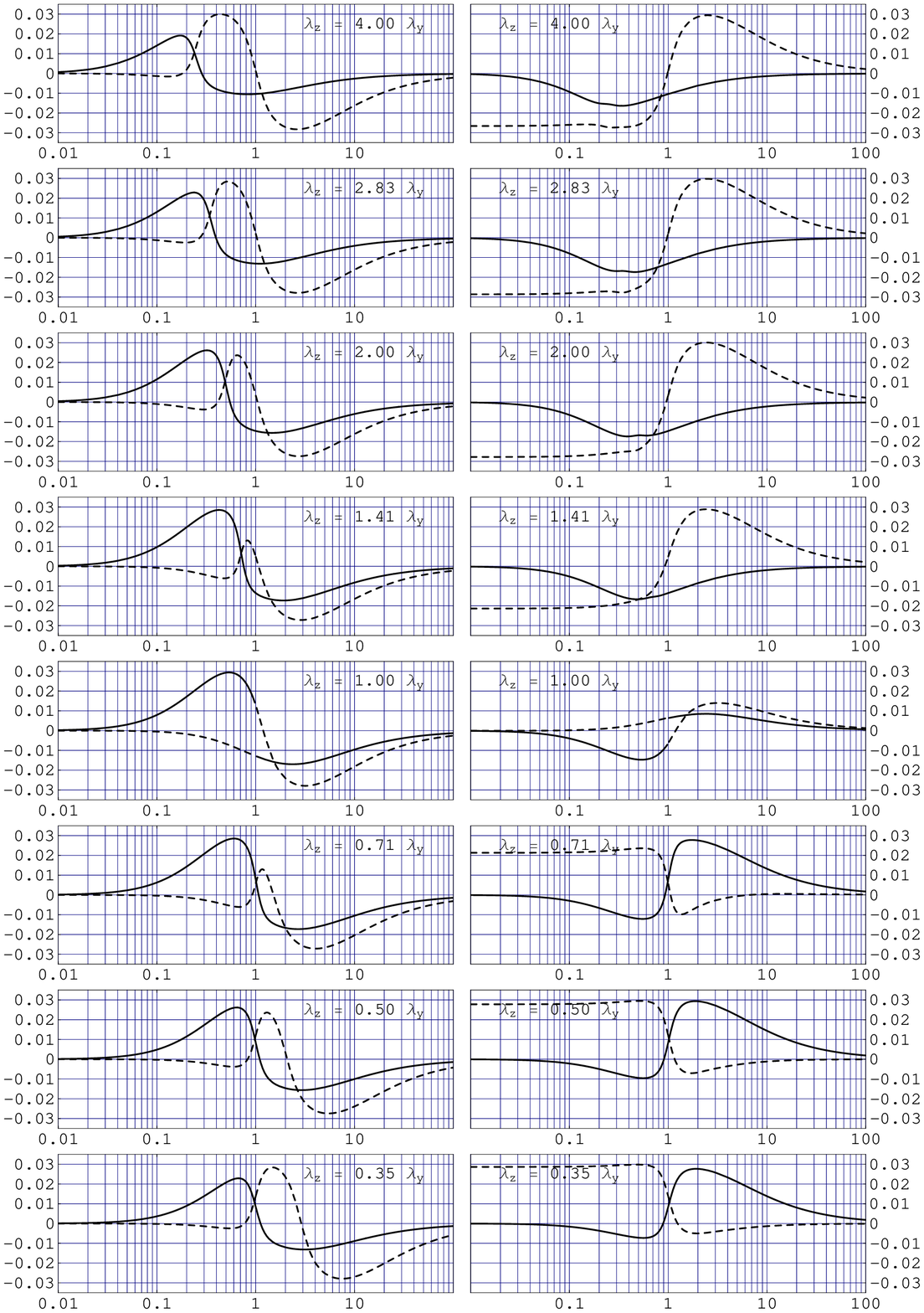} 
\includegraphics[width=25.3mm]{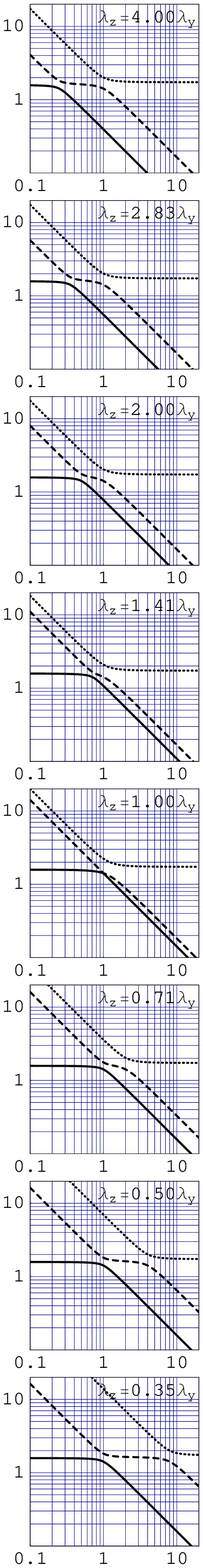} 
\caption{\label{fig2}
Dipolar relative frequency shifts of the collective modes obtained after linear expansion in $\varepsilon_{\mathrm{dd}}$ (for convenience $\varepsilon_{\mathrm{dd}} =0.15$). These shifts depends both on the trap aspect ratios $\lambda_{y}=\omega_{x}/\omega_{y}$, $\lambda_{z}=\omega_{x}/\omega_{z}$ and dipole polarization. The first two columns show the relative frequency shifts $\Delta \Omega / \Omega$ as a function of the logarithm of the trap aspect ratio $\lambda_{y}$. The column on the right shows the Log-Log plot of the unperturbed frequencies as a function $\lambda_{y}$ for reference. 
Dashed lines and solid lines correspond to the intermediate and lower frequency modes, respectively. Different rows of figures correspond to different ratios of $\lambda_{z}/\lambda_{y}$  and polarization. Figures on the left (center) have polarization $\vec{B}||\hat{x}$ ($\vec{B}||\hat{y}$).
From the top $\lambda_{z}/\lambda_{y}=$  $4$, $2.83$, $2$, $1.41$, $1.00$, $0.71$, $ 0.50$, $0.35$. 
Note in the case $\lambda_{z}=\lambda_{y}$, which corresponds to cylindrical symmetry, the dashed and solid lines have a discontinuity in $\lambda_{y}=1$ (that corresponds to the spherical symmetry), due to a level crossing of the two quadrupolar modes. } 
\end{figure*}
\begin{eqnarray}  \label{scaledstatics} 
{\Delta R_{x}^{e} \over \varepsilon_{\mathrm{dd}} R_{x}^{e}\vert_{0}}
&=& -    \frac15 f
-  \frac12 { \omega_{x} \over \omega_y} {\partial f \over \partial \kappa_{1}}
-   \frac12 { \omega_{x} \over \omega_z} {\partial f \over \partial \kappa_{2}}
\label{delta1}\\
{\Delta R_{y}^{e} \over \varepsilon_{\mathrm{dd}}  R_{y}^{e}\vert_{0}} &=&
-    \frac15 f +  \frac12 { \omega_{x} \over \omega_y} {\partial f \over \partial \kappa_{1}}\label{delta2}\\ {\Delta R_{z}^{e} \over \varepsilon_{\mathrm{dd}} R_{z}^{e}\vert_{0}}
&=& -    \frac15 f+   \frac12 { \omega_{x} \over \omega_z} {\partial f \over \partial \kappa_{2}}
\label{delta3} \end{eqnarray}
where $f$ and its partial derivatives $\partial_{\kappa_{1}} \, f$, $\partial_{\kappa_{2}} \, f$ are evaluated in $\kappa_{1}= \omega_{x}/ \omega_y$ and $\kappa_{2}=  \omega_{x} / \omega_y$ for dipole polarized in the $x$-direction. 
Using standard first order perturbation theory in $\varepsilon_{\mathrm{dd}}$ we obtain for the relative frequency shifts 
\begin{equation} \label{nel20}
{\Delta \Omega_{k} \over \Omega_{k}}= {v^{k}_{l}\,\Delta u''_{l m}\,v^{k}_{m}  \over 2 \,(\Omega_{k})^{2}}
\end{equation}
where $v^{k}_{i}$ are the corresponding normalized eigenvectors of the unperturbed matrix (\ref{unpert}). 

Fig. 2 shows the results of the linear approximation in $\varepsilon_{\mathrm{dd}}$. These results allow a quick estimation of the dipolar frequency shift of the two lower collective modes for different geometries.
The upper mode (monopole-like) frequency shifts are not shown in Fig. 2 as they represent a weaker effect of about a factor two. As most of the traps have at least two frequencies that do not differ significantly, we consider the cylindrical symmetry as a starting point (the fifth row in Fig. 2) and we modify the two originally equal trap frequencies by factors ranging between 4 and 0.35. 
We observe that non-axisymmetric traps maximize the frequency difference for different polarization normalized by the unperturbed frequency.
In particular, the relative frequency difference for the intermediate mode reaches a value as large as $0.057$  for $\lambda_{z} \ge 1.41 \lambda_{y}$ and $\lambda_{y}\sim 2$.
\begin{figure} [h!] 
\includegraphics[width=85mm]{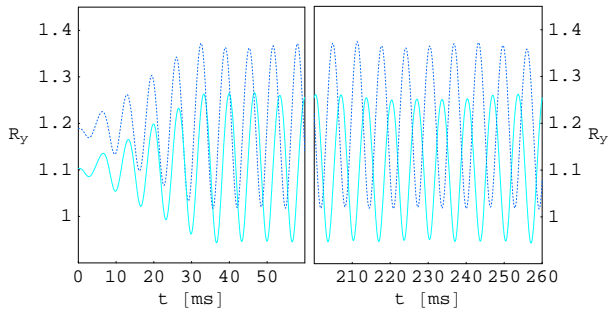} 
\vspace{-0.5cm}
\caption{\label{fig3}
A modulation of $5\%$ of the harmonic trapping potential at frequency of about $\Omega_{2} = 2 \pi 950$ Hz is applied for 5 oscillation periods. The $R_{y}$ radius (arbitrary units) of a chromium BEC ($\varepsilon_{\mathrm{dd}} \approx 0.15 \pm 0.2$) exhibits large oscillations in time of about $50\%$. Solid line and dashed line correspond to the x and y  polarization.
The trap frequencies are assumed $\omega_{x} = 2 \pi 1000$ Hz, $\omega_{y} = 2 \pi 600$ Hz and  $\omega_{z} = 2 \pi 200$ Hz. The modulation frequency is chosen so that the second mode  is excited. The phase shift is quite visible already from the beginning and after 230ms ($\sim 34-36$ periods for the two polarizations) is about $2\times 2\pi$.}
\end{figure}
 
Finally, we discuss how from the measurement of the oscillation frequencies of a condensate for different polarizations we might determine $a$ to the percent level. 
As an example, we consider a chromium BEC confined in a trap with quite distinct frequencies $\omega_{x} = 2 \pi 1000$ Hz, $\omega_{y} = 2 \pi 600$ Hz and  $\omega_{z} = 2 \pi 200$ Hz. In this case, it is possible to excite each collective mode just modulating the laser intensities at approximatively its mode frequency for a few periods of oscillations. Fig. 3 illustrates the process of excitations of the second mode which has a frequency of about $\Omega_{2} = 2 \pi 950$ Hz. The mode oscillation in mainly concentrated along the y-axes (the axes of the middle trap frequency).
Fig. 3 shows the time evolutions of the condensate radius $R_{y}$ for the two different polarization by applying a modulation of $5\%$ of the laser intensities for 5 oscillation periods T. 
We assume that we can see 30-40 clear oscillations. 
When the approximate values of each mode frequency can be inferred from other data we do not need to sample the full time interval and the measurements can be concentrated only at the beginning ($t \sim 35$ ms) and around $t\sim230$ms (about $35$ free oscillations). We estimate that we can measure each frequencies at least with an accuracy of $10^{-3}$. In the case of chromium ($\varepsilon_{\mathrm{dd}} \sim 0.15$), this translates into an accuracy of about $2 \times 10^{-2}$ for the scattering length, which would improve by one order of magnitude the 20$\%$ accuracy of the today's most recent measurements \cite{Griesmaier:2006me}.

{\it Acknowledgments.}
We thank M. Fattori, A. Griesmaier and J. Stuhler for discussions. S.G. thank Professor A. Muramatsu and his group at the Institut f{\"u}r Theoretische Physik III Stuttgart for the hospitality and the technical support.
This work was supported by the Alexander von Humboldt Foundation, the German Science Foundation (DFG) (SFB/TR 21, SFB407 and SPP1116) and the Landesstiftung Baden-W{\"u}rttemberg.

\end{document}